\documentstyle[aps,preprint,eqsecnum,tighten]{revtex}
\begin{document}
\draft
\title{Gravitation vs.\ Rotation in 2+1 Dimensions}
\author{N. J. Cornish}
\address{Department of Physics, University of Toronto, Toronto,
Ontario, Canada M5S 1A7}
\author{N. E. Frankel}
\address{School of Physics, University of Melbourne, Parkville
3052, Australia}
\author{ }

\date{UM-P-93/43, UTPT-93-10}
\maketitle

\begin{abstract}
We investigate rotation and rotating structures in (2+1)-dimensional
Einstein gravity. We show that rotation generally leads to pathological
physical situations.
\end{abstract}

\section*{Introduction}
Rotation in (2+1)-dimensions has recently stimulated much debate
following Gott's\cite{Gott} claim that cosmic strings could act
as time machines by creating closed timelike curves (CTCs) as they pass by
each other. Gott's solution corresponds to point sources
in (2+1)-dimensions with relative angular momentum, and can be mapped
asymptotically to a spinning point source metric. Such spinning point
source solutions contain CTCs and these closed time
circuits are at the heart of the time machine proposal. While it has been
shown\cite{Des} that superluminary velocities are required for Gott's
solution to support timelike lines, the study of rotating, many particle
\cite{clem0} and extended
source solutions has only been partially completed in (2+1)-dimensions
\cite{soleng,willy,clem1}.

Of particular interest are rotating structures since
our previous work\cite{us1,us2} proved that all hydrostatic structures
cause spatial compactification. Since it makes no sense to talk
about such structures rotating, it is clear that any
generalisation of the static results to include rotation will be
non-trivial. This is in sharp contrast to the situation in
(3+1)-dimensions where the solution for a slowly rotating star can
be found by giving a static star a small, uniform spin. Changes to
the pressure and density profiles are found at linear order in
the angular velocity while changes to the structure (oblate flattening)
only occur at second order.

We begin by considering the coordinate systems best suited for studying
stationary solutions to Einstein's equations in (2+1)-dimensions. From
these we recover the spinning point source solution first given by
Deser {\it et al.}\cite{deser}. We then turn to the problem of rotating
structures. In order to simplify the analysis we restrict our attention
to the study of solutions with either shear or vorticity, but not both.

We find that the class of solutions with zero vorticity all encounter
infinite shear while the class of solutions with zero shear all
contain CTCs - unless the structure is supported by tension rather than
pressure. The only solution to avoid some serious pathology is the
rotating annulus with energy density equal to tension. The rotating closed
string solution of Cl\'{e}ment\cite{clem1} is closely related to this
rotating annular structure.

\section{Stationary coordinate systems}
The most general circularly symmetric, stationary metric can be written as
\begin{equation}
ds^{2}=(e^{\nu}dt+\lambda d\theta)^{2}-e^{2\eta}dr^{2}-r^{2}d\theta^{2} \; ,
\label{basic}
\end{equation}
where the metric functions $\nu, \; \eta$ and $\lambda$ are all functions of
$r$ only. Clearly, closed timelike ($ds^2>0$) lines can occur on circles
of radius $r < \lambda$. The vacuum Einstein equations are satisfied if
$\nu, \; \eta$ and $\lambda$ are all constants. Choosing $\nu =0$
as our scaling of time, the other constants can be related to the mass, $M$,
and spin, $S$, of the point source, giving the line element\cite{deser}
\begin{equation}
\label{point}
ds^{2}=dt^{2}+2GS\, dtd\theta-{1 \over (1-GM)^{2}}dr^2-
(r^2-(GS)^{2})d\theta^{2} \; .
\end{equation}
We see that for $r<GS$ the metric admits CTCs and
suffers from having an ill-defined proper spatial volume.

While general coordinate invariance tells us that all coordinate systems
are on an equal footing, operational experience shows us that some are
``more equal'' than others. As an example of this, applying the coordinate
transformations $\tau=t+(GS)\, \theta,\; \rho=(1-GM)^{-1}r$ and
$\phi=(1-GM)\theta$ to the line element in Eqn.\ (\ref{point}) gives
\begin{equation}
ds^{2}=d\tau^{2}-d\rho^{2}-\rho^{2}d\phi^{2} \; . \label{mink}
\end{equation}
All obvious signs that a point source inhabits this spacetime have been
hidden in the coordinates. The time coordinate has become periodic and
the angular coordinate no longer ranges from 0 to $2\pi$. Another example
is the ``fake rotation'' metric
\begin{equation}
ds^{2}={1 \over (1+\Omega_{0}^{2}r^2)}dt^{2}+{2\Omega_{0} r^{2} \over
(1+\Omega_{0}^{2}
r^{2})}dtd\theta-{1 \over (1+\Omega_{0}^{2} r^{2})^{3}}dr^{2}
-{r^2 \over (1+\Omega_{0}^{2} r^{2})}d\theta^{2} \; , \label{fake}
\end{equation}
with $\Omega_{0}=$constant, which is just a fancy way of writing Minkowski
space. This metric arose
in the course of studying rotating structures and corresponds to uniformly
rotating empty space.

To simplify some calculations we will work with the metric
\begin{equation}
ds^{2}=e^{2\kappa}dt^{2}+2\omega\, dtd\theta-e^{2\gamma}dr^{2}
-r^2d\theta^{2} \; , \label{simp}
\end{equation}
rather than the metric in Eqn.(\ref{basic}). In this coordinate
system the spinning point source metric takes the form
\begin{equation}
ds^{2}=dt^{2}+2GS\, dtd\theta-{1 \over (1-GM)^{2}}{1 \over (1+(GS/r)^{2})}
dr^{2}-r^{2}d\theta^{2}\; ,
\end{equation}
so that the region containing CTCs has been mapped onto the region $-(GS)<r<0$.
The singular behaviour of the metric at $r=0$ signals the transition into
the region containing CTCs.

\section{Solving Einstein's equations}

Before solving Einstein's equations, it is instructive to calculate the
expansion, shear and vorticity of the fluid. We are principally interested
in the scalar invariants of these quantities, $\Theta$, $\sigma$ and $\xi$,
respectively, which are defined in terms of the fluids' three velocity,
$u^{\mu}$, as follows
\begin{equation}
\Theta=u^{\alpha}_{;\alpha}\; , \hspace{.4in}
\sigma^{2}={1 \over 2} \sigma^{\alpha\beta}\sigma_{\alpha\beta} \; ,
\hspace{.4in} \xi^2= {1 \over 2}\xi^{\alpha\beta}\xi_{\alpha\beta} \; ,
\end{equation}
where
\begin{eqnarray*}
\sigma^{\alpha\beta}&=&{1 \over 2}(u^{\alpha}_{; \mu}P^{\mu\beta}
+u^{\beta}_{;\mu}P^{\mu\alpha})-{1\over 2}\Theta P^{\alpha\beta} \; ,\\
\xi^{\alpha\beta}&=&{1 \over 2}(u^{\alpha}_{; \mu}P^{\mu\beta}
-u^{\beta}_{;\mu}P^{\mu\alpha}) \; , \\
P^{\alpha\beta}&=&g^{\alpha\beta}-u^{\alpha}u^{\beta} \; .
\end{eqnarray*}
In terms of the metric in Eqn.(\ref{simp}) we find
\begin{eqnarray}
\Theta&=&0 \; , \\
\sigma^{2}&=&(\Omega')^{2}f(r) \; , \\
\xi^{2}&=&(\omega-\Omega r^{2})^{2}g(r) \; ,
\end{eqnarray}
where $\Omega=u^{\theta}/u^{t}$ is the rotational velocity and $f(r)$ and
$g(r)$ are complicated expressions involving the metric functions. The
prime denotes differentiation with respect to $r$.
The above decomposition of $u^{\alpha}_{;\beta}$ allows us to
study solutions with either shear or vorticity, or
both. Obviously any solution with neither shear nor vorticity is
non-rotating.

\subsection{Vorticity Free Solutions}
The vorticity free case is characterised by $\omega=\Omega r^2$. With
this identification the fluid's three velocity takes the form
\begin{eqnarray}
&&u^{t}=(e^{2\kappa}+\Omega^{2}r^{2})^{-{1 \over 2}} \; ,\hspace{.4in}
u^{r}=0 \; , \hspace{.4in}
u^{\theta}=\Omega u^{t} \; , \nonumber \\
&&u_{t}=(e^{2\kappa}+\Omega^{2}r^{2})^{+{1 \over 2}} \; ,\hspace{.4in}
u_{r}=0 \; , \hspace{.4in} u_{\theta}=0 \; .
\end{eqnarray}
The radial component of the energy-momentum conservation equation
$T^{r\alpha}_{\;\; ;\alpha}=0$ gives the hydrostationary equilibrium
condition
\begin{equation}
p'=-{1 \over 2}(\rho +p)\left[\ln(e^{2\kappa}+\Omega^{2} r^{2})\right]'\; .
\end{equation}
The expression for the shear simplifies to
\begin{equation}
\sigma^{2}={(\Omega ' r e^{-\gamma})^{2} \over 2(e^{2\kappa}+
\Omega^{2}r^{2})}\; . \label{shear}
\end{equation}
If we take the limit in which the fluid is essentially a ``test fluid''
(a fluid of test particles) in the field of a point source, we find
\begin{equation}
\sigma^{2}={2G^{2}S^{2}(1-GM)^{2} \over r^{4}} \; .
\end{equation}
The singular nature of the shear at $r=0$ signals that the region below $r=0$
contains CTCs.
We note that the shear vanishes if $S=0$ or $M=1/G$. This is a
useful consistency check since the spinless case should produce no shear
while the limiting case $M=1/G$ corresponds to the cylindrical metric
\begin{equation}
ds^2=dt^2+2GS\, dtd\theta-dr^{2}-a^{2}d\theta^{2} \; , \label{cyl}
\end{equation}
with $a$ an arbitrary constant. The above spacetime is actually non-rotating
since it is spatially compact so again we expect the shear to vanish.
The non-rotating nature of this metric can be seen by implementing a suitable
coordinate transformation or by realising that a test particle on
a rotating cylinder feels no centrifugal force\cite{clem}.

Turning now to the general case, the Einstein equation $G^{t}_{\theta}=0$
gives the condition
\begin{eqnarray}
\label{main}
&&\gamma '(e^{2\kappa}r^2+\omega^{2})(2\omega r-\omega ' r^2)
+(\omega \omega ' +\kappa ' e^{2\kappa}r^{2})(2\omega r -\omega ' r^{2})
\nonumber \\ &&+\omega '' r^2(e^{2\kappa}r^{2}+\omega^{2})-2\omega^{3}
-r^{3}\omega ' e^{2\kappa} =0 \; .
\end{eqnarray}
This equation is trivially satisfied if $\omega=0$ or
$\omega=\Omega_{0}r^{2}$, where $\Omega_{0}$ is a constant, i.e. if the
shear vanishes. Indeed, all
components of Einstein's equations reduce to their non-rotating
expressions for these values of $\omega$. This is because solutions
without shear or vorticity are non-rotating. The unique, non-trivial
solution to Eqn.(\ref{main}) is given by
\begin{equation}
e^{\gamma}={A\, \Omega ' \, r^{3} \over (e^{2\kappa}+\Omega^{2}r^{2})^{1/2}}
\;  \hspace{.4in} (\Omega \neq {\rm constant}), \label{gam}
\end{equation}
where $A$ is a constant of integration. We see that the point source
solution is recovered if $A=1/(2GS(1-GM))$, $\Omega=GS/r^{2}$ and
$\kappa =0$. The point source is the
only example of when the above solution can be used to recover a
static limit. This is because the point source solution does not have
to have $M=1/G$ and so does not necessarily cause spatial
compactification. In all other cases Eqn.(\ref{gam}) is an additional
constraint which does not exist in the non-rotating case $(\Omega=$ constant),
and which is generally incompatible with the static limit.

Further insight can be gained by inserting Eqn.(\ref{gam}) into the
expression for the shear scalar (\ref{shear}), which yields
\begin{equation}
\sigma^{2}={1 \over 2A^{2} r^{4}} \; ,
\end{equation}
thus all vorticity free solutions encounter infinite shear at $r=0$.
In the coordinate system of Eqn.(\ref{basic}) the shear is given by
$\sigma^{2}=1/(2A^{2}[r^2-\lambda(r)^2]^{2})$ so the shear becomes singular at
$r=\lambda(r)$, which marks the boundary of the region containing CTCs.
Even if CTCs are avoided ($r>\lambda(r)$) there will still
be a shear singularity at the origin since $r \rightarrow 0$ then demands
$[r^2-\lambda(r)^2] \rightarrow 0$. Thus all vorticity free solutions are
plagued by infinite shear - with the possible exception of rotating annuli
which do not have matter extending to the origin.

\subsection{Shear Free Solutions}
The shear-free case is characterised by having $\Omega$ constant.
Adopting the metric parameterisation of Eqn.(\ref{basic}) and choosing a
co-rotating frame with:
\begin{eqnarray}
&&u^{t}=e^{-\nu} \; ,\hspace{.6in} u^{r}=0 \; , \hspace{.6in}
u^{\theta}=0 \; , \nonumber \\
&&u_{t}=e^{\nu} \; ,\hspace{.6in} u_{r}=0 \; , \hspace{.6in}
u_{\theta}=\lambda \; ,
\end{eqnarray}
we find that the vorticity scalar is given by
\begin{equation}
\xi = {\lambda \nu ' - \lambda ' \over 2r} e^{-\eta} \; . \label{up}
\end{equation}
The equation $G^{\theta}_{t}=0$ then demands
\begin{equation}
{e^{\nu-\eta} \over r}\left(2\xi \nu' + \xi'\right)=0 \; ,
\end{equation}
so that $\xi=\xi_{0}e^{-2\nu}$. The remaining independent Einstein
equations then read
\begin{eqnarray}
{\eta ' e^{-2\eta} \over r}+3\xi_{0}^{2}\, e^{-4\nu}&=&2\pi G \rho \; , \\
{\nu ' e^{-2\eta} \over r}+\xi_{0}^{2}\, e^{-4\nu}&=&2\pi G p \; ,
\label{a}\\
(\nu '' -\nu ' \eta ' +\nu ' \nu ')e^{-2\eta}
+\xi_{0}^{2}\, e^{-4\nu}&=&2\pi G p \;  . \label{b}
\end{eqnarray}
These can be manipulated to give the equilibrium condition $p' =-(\rho +p)
\nu '$. As first shown in \cite{us2}, Eqns.(\ref{a}) and (\ref{b}) can be
combined to yield
\begin{equation}
{\nu ' \over r}=C e^{\eta -\nu} \; , \label{zip}
\end{equation}
which can then be inserted into Eqn.(\ref{up}) to give
\begin{equation}
\lambda = {\xi_{0} \over C}\left( e^{-\nu}+B e^{\nu} \right) \; .
\label{lamb}
\end{equation}
The arbitrary constant $B$ cannot be determined from Einstein's equations, i.e.
the equations are unaffected by the transformation $\lambda \rightarrow
\lambda + A e^{\nu}$ where A is an arbitrary constant. Such arbitrary terms
can always be removed by a coordinate transformation. The transformation that
achieves this is $t \rightarrow t-(B\xi_{0}/C)\, \theta$. Naturally this
transformation does not affect the geometry, but it does change the topology
of the solution as the time coordinate becomes periodic, just as it did
for the metric in Eqn.(\ref{mink}).

If we demand that the origin be part of our spacetime ($\nu (0)=\eta (0)
=\lambda (0) =0$) then $B=-1$. For this choice of $B$ we find for
small $r$ that $\lambda^2-r^2 =-r^2 (1-\xi_{0}^{2} r^2 +O(r^4))$ so there are
no CTCs near the origin. While the metric is healthy near the
origin, there will generally be CTCs towards the edge of the rotating
structures as the following argument demonstrates.

Following the steps used in \cite{us2}, we can use Eqn.(\ref{zip})
to find a general expression for the pressure in terms of the metric
functions. The solution is
\begin{equation}
p = {1 \over 2 \pi G}\left(C e^{-\eta -\nu} + \xi_{0}^{2}\, e^{-4\nu}
\right) \; . \label{press}
\end{equation}

The above equation shows that all solutions with $C>0$ must be supported
by pressure while solutions with $C<0$ could be supported by either
tension or pressure. However, Eqn.(\ref{zip}) and the equilibrium condition
$p' = -(\rho+p)\nu '$ demand that solutions with $C<0$ are supported by
tension rather than pressure. If this were not the case, the pressure would be
an increasing function of radius which is unacceptable.

In the non-rotating limit ($\xi_{0}=0$) the above expression can be
used to prove that $e^{-\eta} \rightarrow 0$ at the edge of the structure
which in turn proves that all hydrostatic structures cause spatial
compactification.
For $\xi_{0}$ non-zero however, this is no longer true since for $C>0$
the requirement that $p \rightarrow 0$ implies that $e^{-\nu} \rightarrow 0$
so the temporal metric component is singular at the edge of the
structure. Returning to our consideration of CTCs, we see that
the quantity $\lambda^{2}-r^2$ approaches
$[(\xi_{0} /C)^{2}e^{2\nu}-r^{2}]$ at the edge of the structure,
and this quantity is positive for structures with finite radius.
All such structures will thus encounter CTCs. The only solutions with $C>0$ to
escape this pathology
would have pressure profiles which approach zero as $r \rightarrow \infty$,
however these solutions would fill the entire space which takes us back to
the question of a rotating universe. Since the total angular momentum of
a universe must be zero, and since the rotational velocity in this solution
is always of the one sign, we conclude that no solutions of this kind can
exist.

The only hope for finding solutions free of CTCs comes when $C<0$
and the structures are supported by tension rather than pressure.
The structure may then have $p \rightarrow 0$ without
$e^{\nu} \rightarrow \infty$.
It is not surprising that only the tension case can hope to tame
the growth of $e^{\nu}$. Without tension to oppose the centrifugal
force, gravity alone must try and hold the structure together
which in turn leads to the divergence of $e^{\nu}$.
We shall give explicit examples of such solutions in the following section.

\section{Exact Solutions for rotating structures}
It is instructive to consider some explicit examples which serve to illustrate
some of the general results that we have been discussing. We shall begin by
considering some earlier attempts at constructing rotating structures and
relate these solutions to our general picture.

The rotating perfect fluid solution of Williams\cite{willy}
belongs to the class of solutions with zero shear and non-vanishing torsion.
 From our general arguments we expect this solution to contain CTCs for
large $r$ and this is indeed the case.
In the langauge of our paper, the Williams solution has
constant angular velocity $\Omega_{0}= u^{\theta}/u^{t}$, and density,
pressure and vorticity given by
\begin{eqnarray}
\rho = 3p &=& {3\Omega_{0}^{2} \over \pi G (1+\Omega_{0}^{2}r^{2})^{2} } \; ,\\
&& \nonumber \\ \xi &=& {\Omega_{0} \over 1+\Omega_{0}^{2}r^{2} } \; .
\end{eqnarray}
The metric is given by
\begin{equation}
ds^{2}=\left( {1-\Omega_{0}^{2}r^{2} \over 1+\Omega_{0}^{2}r^{2} }
\right)dt^{2}
+\left( {4\Omega_{0} r^{2} \over 1+\Omega_{0}^{2}r^{2} } \right)dtd\theta
-dr^{2}-r^{2}\left({1-\Omega_{0}^{2}r^{2} \over 1+\Omega_{0}^{2}r^{2} } \right)
d\theta^{2} \; .
\end{equation}
This metric encounters CTCs for $r>1/ \Omega_{0}$, and while
the ``gravitational mass'' of the structure is $M_{{\rm G}}=3/G$, a
calculation of the proper mass reveals that $M=(1+i)/G$.

The interior solutions for rotating cosmic strings found by Jensen and
Soleng\cite{soleng} appear to contradict our general results
as they are able to construct solutions supported by pressure that
are free of CTCs. This apparent conflict stems from the fact that they
considered solutions supported by an arbitrarily chosen energy production
mechanism.
In the limit that this heat source is removed, the rotation vanishes and
the solution reduces to empty space. If we had allowed such an {\it ad hoc}
support mechanism in our current work, the additional freedom would have
allowed any choice of metric functions to be a solution to Einstein's
equations.

As has already been mentioned, the closed rotating string solution of
Cl\'{e}ment is an interesting example since it belongs to the
class of solutions where CTCs and shear singularities can be avoided.
Indeed, we can obtain a very similar solution by studying uniformly
rotating tension stars.

For the equation of state $\rho=-\alpha p=\alpha T$, $(\alpha \geq 1)$ the
equation $T'=(\rho-T)\nu'=0$ yields
\begin{equation}
T=T_{0}e^{(\alpha -1)\nu} \; .
\end{equation}
Using Eqns.(\ref{press}) and (\ref{zip}) we find
\begin{equation}
e^{(\alpha +1)\nu}-{\gamma (\alpha +1) \over 2}e^{-2\nu}
=1-{\gamma (\alpha +1) \over 2}-(\alpha +1)(\gamma +1)^2 \pi G T_{0}r^2 \; ,
\end{equation}
where $\gamma = {\xi_{0}}^{2}/(2\pi GT_{0})$. The edge of the structure
is given by $T \rightarrow 0$ which implies $e^{\nu} \rightarrow 0$
and $r \rightarrow \infty$. More precisely, we find
\begin{eqnarray}
e^{-2\nu}&=& 2\pi GT_{0}{(\gamma +1)^2 \over \gamma}r^2+1-{2 \over
\gamma (\alpha +1)}+O(r^{-(\alpha +1)}) \; , \\
\lambda^2 &=& r^2-{2+\gamma(\alpha +1) \over (\alpha +1)(\gamma +1)^2
2\pi G T_{0} }+O(r^{-2}) \; , \\
\lambda e^{\nu} &=& {2 \over (\gamma +1)}\left({\gamma \over 2\pi GT_{0}}
\right)^{1/2} + O(r^{-2}) \; .
\end{eqnarray}
The expression for $\lambda$ shows that the solutions do not encounter
CTCs at large $r$. In fact the metric is actually non-rotating at large $r$
since the gauge transformation
\begin{equation}
\theta \rightarrow \theta -\left[ {\sqrt{\gamma 2 \pi G T_{0}}(\gamma +1)
(\alpha +1) \over 2 + \gamma (\alpha +1) }\right] \, t \; ,
\end{equation}
reduces the metric to a static, cylindrical form. However, we know that this
class of solutions is shear-free, so since the solution is non-rotating at
large $r$ it must be non-rotating everywhere.

There is one exceptional case which avoids this rather disappointing
conclusion. When $\alpha =1$, $T=T_{0}$ and the edge of the structure is
wherever we want it to be. We can even choose to cut the centre out of
such structures as there is no tension gradient to support. The general
solution is given by
\begin{eqnarray}
e^{2\nu}&=&{1 \over 2}\left( a -b r^2 +\sqrt{ 4\gamma +(a -b r^2)^2 }
\right) \; , \\
e^{2\eta}&=&{b\, e^{2\nu} \over 2\pi G T_{0}(4\gamma +(a- b r^2)^2} \; .
\end{eqnarray}
The constants $a$ and $b$ are determined by matching the metric functions
at the inner edge of the annulus, $r=R_{i}$. If we choose $R_{i} \neq 0$
we may then match onto the fake rotation metric (\ref{fake}) and set $B=0$.
This was the choice made by Cl\'{e}ment for the rotating string case. We
can recover Cl\'{e}ment's solution in the limit of vanishing annuluar
width and infinite density (keeping the mass finite as we take the limit).

If we choose instead to match onto ``normal'' Minkowski space we find
$B=-1$, $a=1-\gamma+2\pi GT_{0}(1+\gamma)^2 R_{i}^{2}$ and $b=
2\pi GT_{0}(1+\gamma)^2$. A tedious calculation reveals that the
metric is free of CTCs and that the Euler number of the spatial
manifold can never equal one (spatially compactified) unless $\gamma =0$, in
which case the structure is non-rotating \cite{ft}.

This is most easily demonstrated if we choose $R_{i}=0$ and
$\gamma << 1$. Then the Euler number is given by
\begin{equation}
E=1+\sqrt{1-2\pi GT_{0}R_{0}^{2}}-{\gamma \over 2}(2\pi G T_{0} R_{0}^2)^2
(1-2\pi G T_{0}R_{0}^{2})^{-3/2}+O(\gamma^{2}) \; ,
\end{equation}
and we see that $E > 1$ for any real value of $R_{0}$ (when $\gamma \neq 0$).

Before leaving the rotating annulus or string solutions it is worth
pointing out that the ``rotating dust string'' solution given by Cl\'{e}ment
is actually non-rotating. When the tension vanishes, $\nu '=0$, which in
turn demands that $\xi_{0}=0$ so the string is non-rotating.
This can be seen directly from Cl\'{e}ment's solution as the exterior
metric is given by Eqn.(\ref{cyl}) while the interior metric is
given by Eqn.(\ref{fake}). Both of these metrics actually describe
static rather than stationary solutions.

We conclude our collection of illustrative examples with an example of
our own from the class of solutions with $C>0$. Choosing the equation of state
$\rho =3p$ we find
\begin{eqnarray}
p &=& p_{0} \left( 1-\left({r \over R}\right)^{2} \right)^{2} \; , \\
\xi &=&  \sqrt{2\pi G p_{0} - 1/R^{2} }
\left( 1-\left({r \over R}\right)^{2} \right)^{2} \; ,
\end{eqnarray}
and
\begin{equation}
ds^{2}={1 \over 1-(r / R)^{2}  } dt^{2}
+ {2r^{2}\xi \over (1-(r/ R)^{2})^{3} } dtd\theta
-{ 1 \over ( 1-(r / R)^{2} )^{3} }dr^{2}
-r^{2}\left( {1-2\pi G p_{0} r^{2} \over 1-(r/R)^{2} }\right) d\theta^{2} \; .
\end{equation}
As we expect from our general results, the above metric contains CTCs since
reality of $\xi$ demands $R^{2} \geq 1/(2\pi G p_{0})$ so we see that
the angular coordinate becomes
timelike for $(1/2\pi G p_{0}) < r^{2} < R^{2}$. The CTCs only disappear in
the non-rotating limit, $R^{2} = 1/(2\pi G p_{0})$. A similar exact solution
with $C>0$ can be found for the equation of state $\rho=5p$. This solution
also admits CTCs as expected.

\section{Conclusions}
We have found that the (2+1)-dimensional analogues of rotating stars are
plagued by various pathologies. The vorticity free solutions encounter
infinite shear while the shear free solutions generally encounter CTCs.
It would be interesting to investigate the behaviour of solutions with
both shear and vorticity to see if the combination can (miraculously)
escape the pathologies of the limiting cases. We suspect that the
root of the problem lies in the fact that all these solutions are
spatially compactified in the static limit and this produces a barrier to
rotation.

The only rotating structures to escape these pathologies are tension
annuli. This is consistent with the fact that tension annuli do not
necessarily cause spatial compactification in the static limit.

\section*{Acknowledgments}
We thank Professors G. Cl\'{e}ment and R. Mann for helpful discussions.
One of the authors (NJC) would like to thank the theory group at the
School of Physics, University of Melbourne for their hospitality during
latter part of this work and the support of the Commonwealth Scholarship
Program. This work was funded by the Australian Research Council.


\begin{references}
\bibitem{Gott} J. R. Gott III, Phys. Rev. D{\bf 46}, 1126 (1991).
\bibitem{Des} S. Deser, R. Jackiw and G. 't Hooft, Phys. Rev. Lett.
{\bf 68}, 2647, (1992); see also Ann. Phys. (N.Y.) {\bf 152}, 220 (1984).
\bibitem{clem0} G. Cl\'{e}ment, Int. J. Theor. Phys. {\bf 23}, 335, (1984).
\bibitem{soleng} H. Soleng, Gen. Rel. Grav. {\bf 24}, 1131, (1992);
B. Jensen and H. Soleng, Phys. Rev. D{\bf 45}, 3528, (1992).
\bibitem{willy} J. Williams, Gen. Rel. Grav. {\bf 23}, 181 (1991).
\bibitem{clem1} G. Cl\'{e}ment, Ann. Phys. (N.Y.) {\bf 201}, 241, (1990).
\bibitem{us1} N. J. Cornish and N. E. Frankel, Phys. Rev. D{\bf 43},
2555 (1991).
\bibitem{us2} N. J. Cornish and N. E. Frankel, Phys. Rev. D{\bf 47},
714 (1993).
\bibitem{deser} S. Deser, R. Jackiw, and G. 't Hooft, Ann. Phys. (N.Y.),
{\bf 152}, 220, (1984).
\bibitem{clem} G. Cl\'{e}ment, Class. Quantum Grav. {\bf 9}, 2615, (1992).
\bibitem{ft} In our earlier treatment of the tension annulus \cite{us2}
we incorrectly stated that $E=GM=1$. This does not have to be the case.
\end{references}
\end{document}